\documentclass{elsart}

\usepackage{natbib}

\usepackage{graphics}

\usepackage{amssymb}
\def\lsim{\lower 2pt \hbox{$\, \buildrel {\scriptstyle <}\over
         {\scriptstyle \sim}\,$}}
\def\gsim{\lower 2pt \hbox{$\, \buildrel {\scriptstyle >}\over
         {\scriptstyle \sim}\,$}}

\begin{document}

\begin{frontmatter}



\title{Pulsars and Their Nebulae as $EGRET$ Sources}

\author[label1,label2]{Mallory S.E. Roberts}
\address[label1]{McGill Universtity, 3600 University, Montr\'eal, QC, H3A 2T8, Canada}
\address[label2]{Eureka Scientific, 2452 Delmer St., Ste. 100, Oakland, CA 94602, USA }
\ead{roberts@physics.mcgill.ca}
\ead[url]{http://www.physics.mcgill.ca/$\sim$roberts}


\begin{abstract}

At the end of the $EGRET$ mission, only 6-8 Galactic sources 
had been identified as young pulsars. Since then, several 
energetic pulsars have been discovered in $EGRET$ error boxes 
along the Galactic plane,  as well as  several pulsar wind nebulae 
from which pulsations have not yet been discovered. Some of these 
nebulae are associated with moderately variable $EGRET$ sources, suggesting 
that the $\gamma$-ray emission might be coming from the nebula 
rather than from the pulsar magnetosphere. There is also a population 
of unidentified $EGRET$ sources at mid-Galactic latitudes which have been 
proposed to be either nearby middle-aged pulsars or millisecond pulsars. 
I review the current status of observational 
studies of pulsars associated with $EGRET$ sources and what they 
suggest the upcoming $AGILE$ and $GLAST$ $\gamma$-ray missions might observe.

\end{abstract}


\end{frontmatter}

\section{The $EGRET$ Legacy}
\label{leg}

The $EGRET$ instrument (1991-1999) on board the Compton Gamma-Ray Observatory 
produced the best data currently available on the 30 MeV - 20 GeV sky.  
It detected nearly 300 \citep{hbb+99,lm97} 
discrete sources of high-energy $\gamma$-rays. The $EGRET$ sources
seem to have at least three spatial distributions. There is a nearly
isotropic population, nearly all ($\sim$ 100) of which have now been 
plausibly associated with blazars, a radio bright sub-class of active 
galactic nuclei \citep{srm03,srmu04,mhr01}. There is a population of
bright sources with generally hard spectra within a few degrees
of the Galactic plane. 6-8 of these have been identified as young, energetic
pulsars by direct detection of $\gamma$-ray pulsations at periods previously
known through radio or X-ray observations \citep{klm+00,rfh+96,nfl+96}.
These pulsars all have fairly hard $\gamma$-ray spectra (photon power-law 
index $\Gamma \lsim 2.2$) and their emission averaged over many pulsations
is very steady. The existence of Geminga, a nearby $\gamma$-ray pulsar 
from which there has been no confirmed detection of radio pulsations despite
many extensive searches, and the spatial coincidence of unidentified
$EGRET$ sources with regions of recent star formation \citep{kc96,yr97,rbt99} 
suggest
many of the unidentified sources could be young pulsars not previously detected,
or even undetectable, in radio. However, the log N-log S distribution of the
unidentified sources is not the same as that of young pulsars \citep{bam+04},
and many of the sources show evidence for variability \citep{mmct96,ntgm03}, 
indicating more than one source class.

There is a third population of sources at mid-Galactic latitudes
($5^{\circ} \lsim b \lsim 30^{\circ}$) which is fainter and which generally have steeper 
spectra than the low-latitude sources. The spatial distribution is similar to that of the 
collection of nearby regions of recent star formation known as the Gould belt, with possibly the 
addition of a halo population \citep{gre01}. However, it is also similar to the distribution
of millisecond pulsars \citep{rom01} and there was one marginal detection
of pulsed emission from the millisecond pulsar PSR J0218+4232 
in the $EGRET$ data \citep{khv+00}.

Here I will discuss observational progress that has been made since the demise
of CGRO in the study of pulsars as sources of the emission detected by $EGRET$.  
These studies are preparing the ground for future $\gamma$-ray missions such
as $AGILE$ and $GLAST$. For a recent review of high-energy emission from pulsars
and their nebulae, see \citet{krh04}.

\section{High Resolution X-ray Imaging of $\gamma$-Ray Pulsars: Constraining the 
Geometry}

Energetic pulsars accelerate particles to high energies in their magnetospheres.
These particles can interact with the surrounding medium to produce synchrotron nebulae
which are bright in radio and X-rays 
(For references to individual nebulae, see the on-line Pulsar
Wind Nebula (PWN) Catalog at http://www.physics.mcgill.ca/$\sim$pulsar/pwncat.html).
Around the youngest, most energetic sources, the radio emitting parts of these nebulae tend to be
rather amorphous, but the X-ray emitting regions can be highly structured. 
The high spatial resolution of the $Chandra$ satellite has allowed these
structures to be resolved. The three youngest (characteristic age $\tau < 20,000$ yr)
$\gamma$-ray emitting pulsars, Crab, Vela, 
and PSR B1706$-$44, all have toroidal nebulae with perpendicular jets. 
The tori are presumably produced by equatorial winds, while the jets are aligned with the 
spin axes. The middle aged pulsar ($\tau\sim 100,000$ yr) PSR B1951+32 has a RPWN 
which is distorted by the pulsar's motion, while the two oldest pulsars ($\tau > 300,000$ yr), 
PSR B1055$-$52 and Geminga, have very faint PWN. 

The location of the particle accelerating region in the magnetosphere is still being debated 
\citep[see eg.][and elsewhere in these proceedings]{dh96,rom96a,hm98}.
The polar cap class of models presume the emission is from a region near the magnetic poles,
the outer gap class of models assume the emission comes from near the light cylinder, 
while the more recent slot gap models bridge the two regions. All of the models in principle
can predict the shape of the $\gamma$-ray pulse profile, given the overall geometry of the pulsar: i.e.
the viewing angle between the spin axis and the observer ($\zeta$) and the magnetic inclination angle 
between the spin axis and the 
magnetic axis. The outer gap models have been fairly successful in reproducing 
the observed $\gamma$-ray pulse profiles \citep{ry95}, but the geometry was, 
in general, unconstrained. \citet{dhr04} compare $\gamma$-ray pulse 
and optical polarization profiles predicted by the three models for the
Crab and Vela. The three models best fits to the $\gamma$-ray light curves
all require different $\zeta$ values.

The new observations of toroidal PWN put powerful constraints on the 
pulsar geometry. By only assuming the torus is symmetric about the equator, 
the ellipticity of the observed torus and the brightness asymmetry due to
Doppler boosting immediately give the three dimensional orientation 
of the spin axis of the pulsar \citep{rn03,nr04}. In addition, there
may be a relationship between the thickness of the torus (or separation of double torii) to the magnetic
inclination angle. Since the radio pulse polarization sweep also is thought to 
give information about the pulsar geometry \citep{rc69a}, the geometries
of some pulsars are now quite strongly constrained. 
Unfortunately, only Crab, Vela and PSR B1706$-$44 of the known 
$\gamma$-ray pulsars have PWN useful for this type of fitting. 

Another interesting development in the post-EGRET era has been the
better characterization of hard X-ray pulse profiles with the Rossi X-ray 
Timing Explorer (RXTE). Hard X-rays are thought to arise from synchrotron 
radiation emitting particles which are a by-product of the acceleration and
emission of the primary particles responsible for the $\gamma$-ray emission. 
Thus, the hard X-ray pulse profiles should be intimately related to the 
$\gamma$-ray profiles. Studies of the Vela pulsar show it to
have a quite complicated collection of pulse peaks \citep{hsg+02} whose phases 
and spectra need to be reconciled with the geometrical constraints from
its nebula and the $\gamma$-ray pulsations. 

\section{Potential New $EGRET$ Pulsars}

The ubiquity of X-ray PWN around the most energetic pulsars allows 
the discovery of new potential $\gamma$-ray emitting pulsars through
X-ray imaging. 2-10 keV imaging with the $ASCA$ satellite 
of $EGRET$ sources with significant GeV emission revealed several 
interesting new potential PWN \citep{rrk01,hglh01}. The X-ray localizations
allowed for very deep radio pulse searches of these sources, two of
which were discovered to contain the energetic but faint ($< 1$mJy) radio
pulsars PSR J2229+6114 \citep{hcg+01} and PSR J2021+3651 \citep{rhr+02}.
High-resolution imaging with Chandra revealed that both 
have PWN which are probably toroidal 
and hence useful geometrical constraints can be made \citep{nr04,hrr+04}. 
These are particularly exciting since $\gamma$-ray pulsations 
are almost certain to be discovered by the upcoming $AGILE$ and $GLAST$
missions (searching the $EGRET$ data is problematic since extrapolation of
the pulse ephemeris back several years is complicated by timing noise and 
pulse glitches in young pulsars such as these). 
The geometrical constraints allow for the prediction of the $\gamma$-ray 
pulse shape and phase. 

A major recent development in pulsar science has been the completion of the
Parkes Multibeam Survey of the Galactic plane, which has doubled the
population of known radio pulsars \citep{mlc+01}. Several of the pulsars 
found in this survey are energetic young pulsars coincident with $EGRET$ 
error boxes \citep{dkm+01,cbm+01,kbm+03}. The three multibeam pulsars most 
likely to be counterparts of the $EGRET$ sources have all been imaged with
Chandra. The most energetic one, PSR J1420$-$6048, has a faint X-ray 
nebula with just a hint of structure in the short ($\sim 10$ ks) exposure
Chandra image. It is contained within a non-thermal radio nebula that is
part of the Kookaburra radio complex \citep{rrjg99,rrj01}. PSR J1016$-$5857
also has a very faint nebula with uncertain structure \citep{cgg+04}. It is unclear if either PWN will be useful 
for obtaining geometrical constraints. The third pulsar, PSR J1837$-$0604, 
was not detected by a short Chandra observation.

Two other hard, GeV emitting sources are coincident with X-ray sources that are likely 
radio-quiet pulsars. 3EG J1835+5918 is the only unidentified high Galactic
latitude source which is bright above 1 GeV \citep{lm97} and has very 
steady emission. There is a dim
X-ray source with no optical counterpart, strongly suggesting
it is a neutron star \citep{hgmc02,rbc+01}.  All other X-ray and radio sources in 
the error box have been ruled out as counterparts. This is 
a strong case for a relatively old and nearby $\gamma$-ray pulsar
like Geminga. In CTA 1, an X-ray PWN was discovered with $ASCA$ with 
an associated ROSAT point source 
\citep{ssb+97,brkc98}. A recent Chandra image shows 
an unusual bent jet-like structure \citep{szh+04,hgc+04}.
In both cases, very deep radio searches have failed to detect
pulsations from the apparent neutron stars.

\begin{table*}[h!]
{\small
\caption[]{Known and Potential $EGRET$ Pulsars and PWN}
\vspace{0.2cm}
\label{pwntab}
\begin{tabular}{lccccccc}
\hline
Name & 3EG Name & Type$^a$ & log$\dot E$ & $\Gamma^b$ & $\zeta^c$ & $V_{12}^d$ & $\delta^d$ \\
\hline
CTA 1 & 3EG J0010+7309  & ? & -- &  $1.85\pm.10$ & -- & 0.40 & $0.26^{+.47}_{-.26}$ \\
PSR J0218+4232 & 3EG J0222+4253 & M & 35.4 & $2.6^e$ & -- & -- & $0.0^{+.37}_{-.00}$ \\
Crab & 3EG J0534+2200  & S & 38.7 & $2.19\pm.02$ &  $62^{\circ}$ & -- & $0.08^{+.07}_{.05}$  \\
Geminga & 3EG J0633+1751  & R & 34.5 & $1.66\pm .01$ & --  & -- & $0.10^{+.06}_{.04}$\\
Vela & 3EG J0834$-$4511  & S & 36.8 & $1.69\pm .01$ & $64^{\circ}$& 0.61 & $0.16^{+.12}_{-.06}$ \\
PSR J1016$-$5857 & 3EG J1013$-$5915  & ? & 36.4 & $2.32\pm.13$ & -- & 0.18 & $0.14^{+.38}_{-.14}$ \\
PSR B1046$-$58 & 3EG J1048$-$5840  & ? & 36.3 & $1.97\pm .09$ & -- & -- & $0.0^{+.31}_{-.00}$ \\
PSR B1055$-$52 & 3EG J1058$-$5234 & ? & 34.5 & $1.94\pm .10 $ & -- & -- & $0.0^{+0.47}_{-.00}$ \\
PSR J1420$-$6048 & 3EG J1420$-$6038  & ? & 37.0 & $2.02\pm .14$ &  -- & 1.59 & $1.03^{+.80}_{-.66}$\\
Rabbit & 3EG J1420$-$6038  & R & -- & $2.02\pm .14$ &  -- & 1.59 & $1.03^{+.80}_{-.66}$ \\
PSR J1614$-$2230 & 3EG J1616$-$2221 & M & 34.1 & $2.42\pm .24$ & -- & -- & $0.0^{+.54}_{-.00} $ \\ 
PSR B1706$-$44 & 3EG J1710$-$4439 & S & 36.5 & $1.86\pm .04$ & $55^{\circ}$ & -- & $0.07^{+.14}_{-.07}$ \\
G359.89$-$0.08 & 3EG J1746$-$2851  & R & -- & $1.70\pm .07$ & -- & 2.35 & $0.48^{+.27}_{-.19}$ \\
G7.4$-$2.0 & 3EG J1809$-$2328  & R & -- & $2.06\pm .08$ & -- & 3.93 & $0.71^{+.42}_{-.25}$\\
G18.5$-$0.4 & 3EG J1826$-$1302  & R & -- & $2.00\pm .11$ & -- & 3.22 & $0.88^{+.57}_{-.38}$ \\
RX J1836.2+5925 & 3EG J1835+5918 & G & -- & $1.69\pm .07$ & -- & 0.09 & $0.15^{+.32}_{-.15}$ \\ 
PSR J1837$-$0604 & 3EG J1837$-$0606 & -- & 36.3 & $1.82\pm .14$ & -- & -- & $0.0^{+.67}_{-.00}$ \\
PSR B1853+01 & 3EG J1856+0114  & R & 35.6 & $1.93\pm .10$ & -- & 1.57 & $0.71^{+.82}_{-.43}$\\
PSR B1951+32 & --$^f$ & R & 36.6 & $1.7\pm .1$ & -- & -- & -- \\
PSR J2021+3651 & 3EG J2021+3716 & S & 36.5 & $1.86\pm .10$ & $83^{\circ}$ & 0.71 & $0.36^{+.47}_{-.33}$ \\
PSR J2229+6114 & 3EG J2227+6122 & S & 37.4 & $2.24\pm .14$ & $46^{\circ}$ & 0.21 & $0.20^{+.62}_{-.20}$\\
\hline
\end{tabular}
\vskip 0.3cm
$^a$ R=RPWN, S=SPWN, ?=PWN type unclear, G=Possible Gould Belt, M=MSP\\
$^b$ $EGRET$ photon spectral index \citep[from][save where noted]{hbb+99}\\
$^c$ spin axis inclination angle \citep{nr04,hrr+04}\\
$^d$ variability indices from \citep{ntgm03} -- see text\\
$^e$ estimated contribution from pulsar \citep[see][for details]{khv+00}\\
$^f$ Only seen in pulsations \citep{rfh+96}\\
}
\vspace{-1cm}
\end{table*}
\section{RPWN as Variable $EGRET$ Sources}

While several of the hard, non-variable sources in the Galactic plane now have potential 
pulsar counterparts, what about the variable sources? Are the $\gamma$-ray emitting particles accelerated
by some other type of engine, or do pulsars play a role with these sources as well? Observationally, 
it seems like many of these sources probably are associated with young pulsars, but ones in a special situation. 
The $ASCA$ X-ray survey of the brightest sources of GeV emission by \citet{rrk01} identified four nebula
coincident with apparently variable sources. Subsequent radio and high-resolution X-ray imaging has
shown that all of them are rapidly-moving, ram-pressure confined pulsar wind nebulae (RPWN) 
\citep{rbg+04}. 

The advent of $Chandra$ has allowed the determination of the X-ray structure of
RPWN. The Mouse nebula, towards the Galactic centre, is one of the brightest RPWN
in X-rays and promises to become the canonical example. In radio, it consists of a bright head and 
shoulders on a fainter body with a long tail trailing behind. The X-ray emission is mostly confined
to the head, and consists of a point source at one end of an elliptical region of bright emission 
(the `tongue') extending back towards
the radio body, with a fainter halo around it \citep{gvc+03}. Other sources tend to be fainter in X-rays,
and generally just a narrow, elongated region is evident. 

The most detailed study of $EGRET$ source variability is that of \citet{ntgm03}. They attempt to
answer both how variable a source is ($\delta = \sigma_F/\mu_F$) and 
whether or not it is variable at all (probability for rejecting non-variable hypothesis $P=(1-10^{-V_{12}})$). 
They list only two sources with $V_{12} > 3.0$ that have not been associated with Active Galactic Nuclei, both of
which are at low Galactic latitudes. 
The amplitude of the variability is moderate ($\delta \sim 0.8$), and, since the flux points of the study were by
viewing periods which were typically of a few week duration, the timescale of the variability is on the the order of a few months to a few years. 

The X-ray nebula in GeV J1809$-$2327/3EG J1809$-$2328 was discovered in the $ASCA$ survey and subsequent VLA imaging revealed 
a funnel shaped radio PWN \citep{rrk+01}. Chandra ACIS imaging \citep{brrk02} and a small window XMM-Newton PN image both
show a point X-ray source near a complex of X-ray emitting massive stars.
This is at
the tip of the radio nebula and there is an X-ray  trail leading back towards
the nebula's center. There is also a larger,
fainter radio nebula which may be similar the the body of the Mouse. The PWN 
seems to have come from an area of soft X-ray 
emission seen by ROSAT (G7.4$-$1.4) that may be a thermal composite supernova remnant.
 
The region of the Galactic plane near $l=18^{\circ}$ contains an unidentified source of bright MeV emission
discovered by COMPTEL and at least two $EGRET$ sources. 90~cm imaging of the area shows several SNR and numerous 
other possible non-thermal sources (Brogan et al. in preparation). GeV J1825$-$1310 / 3EG J1826$-$1302 contains a complex of
X-ray emitting massive stars near which is another X-ray point source with a non-thermal trail of emission. There is a hint
of a larger bow-shock shaped nebula in the short Chandra image, which is consistent with the larger emission area in the
ASCA image. The 90~cm image shows a possibly related radio structure. 

There are two other $EGRET$ sources with $V_{12} > 1.5$ which contain apparent RPWN. PSR B1853+01 in the X-ray composite
supernova remnant W44 has a faint radio and X-ray RPWN \citep{pks02,fggd96}. 
The Kookaburra complex contains two PWN, the Rabbit nebula and one around PSR J1420$-$6048 \citep{rrjg99}. Recent 
Chandra and XMM-Newton observations suggest the Rabbit is an RPWN, but with significant X-ray
emission not confined by the apparent bow-shock. 
The radio source G359.89$-$0.08, coincident with the $EGRET$ source near the
Galactic center, is also a possible RPWN \citep{lwl03}. All told, about half the low-latitude $EGRET$ sources with good evidence for variability 
are known to contain RPWN. 

Currently, there is no developed theory for variable GeV emission from RPWN. However, there is adequate energy in the 
wind, and the magnetic fields are strong enough for rapid synchrotron cooling of $\gamma$-ray emitting electrons. 
There is a theory for the variability of the X-ray and optical emitting ``wisps" in the Crab nebula which
might be adapted for these sources \citep{sa04}. It should also be noted that the motion of the pulsar through
regions of varying density gives another natural mechanism for the variability.

\section{Mid-Latitude Sources as Pulsars}

Since pulsars can successfully account for many of the Galactic plane sources, is 
it possible that they can make up a substantial portion of the mid-latitude sources?
Two arguments against this are that their spectra tend to be significantly softer than the 
known $\gamma$-ray pulsars, and that many of them appear to be variable. However, estimates
of the supernova rate in the Gould belt suggest a fair number of nearby pulsars should have been born
in the last few million years. In either the outer-gap \citep{czlj04} or polar-cap model \citep{hz01}, 
a few dozen should be detectable
in $\gamma$-rays, with most being seen at relatively large angles to the radio beam. Hence, most should
be radio faint. Furthermore, in the polar-cap model, the faint, outer beam of $\gamma$-ray emission should be softer
than the brighter, core emission. Some of the pulsars should be bright and 
hard; perhaps this is accounted for by 3EG J1835+5918 and Geminga.
The case for possible
millisecond pulsar counterparts is less well developed, but the one claimed detection
suggests they may also have somewhat steeper spectra.

Practically speaking, the mid-latitude sources are difficult for counterpart searches because they have large error
boxes (95\% confidence contours $\sim 1.5^{\circ}$ in diameter). Gould belt pulsars should be mostly radio
quiet, although some should have detectable radio emission. While it is hoped that blind searches for
pulsations will be possible with $GLAST$, a known radio or X-ray pulse ephemerides was used for the discovery
of all the confirmed $\gamma$-ray pulsars. For millisecond pulsars, since the parameter search space is much greater 
and they tend to be in binary systems, a known ephemeris is crucial for $\gamma$-ray pulse searches. 

It is therefore desirable to find possible $\gamma$-ray emitting radio pulsars. However, a recent search 
of 56 mid-latitude error boxes with the Parkes telesope at 20~cm \citep{rrh+03} 
detected no potential Gould belt pulsars and only one marginal possibility (PSR J1614$-$2230) 
for a $\gamma$-ray millisecond pulsar. If a majority of the sources are pulsars, one would have expected a few 
more reasonable candidates. One caveat of the search is that the Gould belt pulsars are expected to be very close
and hence have very low dispersion measures, making the relatively long period pulsars hard to distinguish 
from local radio frequency interference. How this limited the survey sensitivity is hard to quantify, 
but broad-band, low-frequency searches would be better suited for detecting such pulsars. 

\section{Conclusions}

Around half of the unidentified sources that are bright above 1 GeV now have plausible associations with 
young pulsars, and several of the other Galactic $EGRET$ sources do as well. High resolution X-ray images
are helping to constrain the geometries of some of these pulsars, but in most cases deeper observations are needed to confirm the results. 
There is growing evidence that a substantial fraction of the bright, variable Galactic plane sources
are associated with RPWN, but the X-ray data is poor due to short exposures and a robust theory of 
GeV emission from RPWN is still lacking. While it is plausible that many of the unidentified sources at
mid-Galactic latitudes are pulsars, the lack of plausible radio counterparts suggest that most of these
sources are not pulsars. However, other source classes may be harder to identify, and so exhaustive
pulse searches are still called for. $AGILE$ should be able to detect pulsations
from any source detected by $EGRET$ if there is a current pulse ephemerides.
It should also better localize the bright variable sources. $GLAST$ will
obtain excellent light curves and positions for the variable sources allowing
for correlation studies with observations at lower energies.




\bibliographystyle{apj}

\bibliography{journals_apj,egret,malloryrefs,modrefs,psrrefs,crossrefs}




\end{document}